\documentclass[twocolumn]{aastex631}

\usepackage{amsmath}

\newcommand{\rr}[1]{{\textcolor{black}{#1}}}

\begin{document}

\title{Are long gamma-ray bursts progenitors to merging binary black holes?}

\author{Tom Y. Wu}
\affiliation{David A. Dunlap Department of Astronomy and Astrophysics, University of Toronto, 50 St George St, Toronto ON M5S 3H4, Canada}

\author[0000-0002-1980-5293]{Maya Fishbach}
\affiliation{Canadian Institute for Theoretical Astrophysics, 60 St George St, University of Toronto, Toronto, ON M5S 3H8, Canada}
\affiliation{David A. Dunlap Department of Astronomy and Astrophysics, University of Toronto, 50 St George St, Toronto ON M5S 3H4, Canada}
\affiliation{Department of Physics, 60 St George St, University of Toronto, Toronto, ON M5S 3H8, Canada}

\begin{abstract}
The distribution of delay times between the formation of binary black hole (BBH) progenitors and their gravitational-wave (GW) merger provides important clues about their unknown formation histories. When inferring the delay time distribution, it is typically assumed that BBH progenitor formation traces the star formation rate (SFR). 
In this work, we consider the rate of long gamma-ray bursts (LGRBs) instead of the SFR.
LGRBs are thought to correspond to the formation of (possibly spinning) black holes, and may therefore be related to the BBH progenitor population. 
By comparing the redshift evolution of the LGRB rate as inferred by~\citet{2022ApJ...932...10G} and the BBH merger rate inferred by LIGO-Virgo-KAGRA (LVK) observations, we find that the delay time distribution between LGRBs and BBH mergers is well-described by a power law with minimum delay time $10$ Myr and slope $\alpha ={-0.96}^{+0.64}_{-0.76}$ (90\% credibility). This matches theoretical expectations for the BBH delay time distribution, which in turn lends support to the hypothesis that LGRBs trace BBH progenitor formation. 
However, comparing the absolute rates of these two populations, we find that at most $f = {4}^{+10}_{-2}\%$ of LGRBs may evolve into merging BBH.
We also consider the possibility that LGRBs only produce BBH systems with large aligned spins (with effective inspiral spin $\chi_\mathrm{eff} > 0.2$). In this case, we find $f = 0.3^{+1.0}_{-0.2}\%$ and the delay time distribution favors the steepest power-law slopes we consider ($\alpha = -2$).
We argue that asynchronous observations of LGRBs and GWs provide a powerful multimessenger probe of black hole lifecycles across cosmic history. 
\end{abstract}

\section{Introduction}
\label{sec:intro}
There are many proposed evolutionary channels for making merging binary black hole (BBH) systems \citep[see, e.g.,][for reviews]{2020FrASS...7...38M,2022PhR...955....1M}. Merging BBH systems likely evolve from massive stars that collapse into black holes (BHs) at the end of their lives. Under rare conditions, binary star evolution and/or stellar dynamics cause a BH stellar remnant to end up in a tight binary with another BH. If the binary is sufficiently tight, its orbit decays due to gravitational-wave (GW) radiation and the two BHs inspiral towards each other and finally merge into one. It often takes billions of years for two BHs to merge due to GW radiation, because the inspiral time $\tau_\mathrm{insp}$ strongly depends on the initial orbital separation $a$; for circular binaries, $\tau_\mathrm{insp} \propto a^4$~\citep{1964PhRv..136.1224P}. Because massive stars die so quickly ($\lesssim 10$ Myr), the delay time $\tau$ between the formation of the progenitor stars and the merger of the BBH system is almost entirely dominated by the GW inspiral phase, $\tau \approx \tau_\mathrm{insp}$.
This delay time depends on the initial binary orbit, and thus on the astrophysical processes responsible for creating the BBH system. 

Many formation channels predict delay time distributions that are well-approximated as a power law $p(\tau) \propto \tau^\alpha$ with slopes $-1.5 \lesssim \alpha \lesssim -0.3$ above a minimum delay time of $\tau_\mathrm{min} \sim 0.1$ Gyr~\citep[e.g.][]{2010ApJ...716..615O,2012ApJ...759...52D,2018ApJ...866L...5R,2019MNRAS.487....2M,2020MNRAS.498..495D,2023ApJ...957L..31F,2024ApJ...967...62Y,2024arXiv240501623B}. \rr{Typical population synthesis simulations of isolated binary evolution via a combination of stable mass transfer and common envelope~\citep{2010ApJ...716..615O,2012ApJ...759...52D,2023ApJ...957L..31F}, dynamical assembly in dense star clusters~\citep{2024ApJ...967...62Y}, and BBH mergers from young star clusters~\citep{2020MNRAS.498..495D} predict delay time distributions with $\alpha \approx -1$. Other formation channels, like stellar triples~\citep{2017ApJ...841...77A} or chemically homogeneous evolution~\citep{2016MNRAS.458.2634M,2020MNRAS.499.5941D}, typically predict relatively longer delay times, $\alpha > -1$. Because GW radiation is inefficient at shrinking the binary orbit, delay time distributions steeper than $\alpha \lesssim -1$ are rarely predicted from any formation scenario.}

In the past decade, the GW detector network consisting of LIGO, Virgo and KAGRA~\citep{2015CQGra..32g4001L,2015CQGra..32b4001A,2021PTEP.2021eA101A} has observed the GW radiation from $\mathcal{O}(100)$ BBH mergers. The third Gravitational-Wave Transient Catalog GWTC-3 \citep{2023PhRvX..13d1039A} includes $\approx 70$ confident BBHs (with false-alarm rate FAR $<$ 1 per year). Previous studies~\citep{2021ApJ...914L..30F,2024ApJ...972..157V,2024ApJ...967..142T,2024ApJ...970..128S} used these events to infer the BBH delay time distribution by comparing the redshift evolution of the BBH merger rate to the star formation rate (SFR). This assumes that the progenitors of BBH mergers trace the SFR. 
However, BBH progenitors are likely biased tracers of star formation because the fraction of stars that produce merging BBH systems depends on conditions that vary over cosmic time, like the metallicity, the stellar initial mass function, and the binary fraction~\citep[e.g.][]{2010ApJ...715L.138B,2018A&A...619A..77K,2021MNRAS.502.4877S,2024AnP...53600170C,2024arXiv240501623B}. Earlier studies that measured the BBH delay time distribution therefore included a metallicity-dependent formation efficiency, effectively assuming that BBH progenitors trace the low-metallicity SFR. 
This introduces additional uncertainty because the low-metallicity SFR is not well known~\citep[see, e.g.][for a review of the effect on GW populations]{2024AnP...53600170C}. 
Nevertheless, assuming reasonable values of the metallicity-specific SFR from observations~\citep[e.g.][]{2017ApJ...840...39M} and the metallicity-dependent BBH formation efficiency from theoretical models~\citep[e.g.][]{2013ApJ...779...72D}, the BBH delay time distribution was inferred to strongly favor short delay times (power law slopes $\alpha < -1$;~\citealt{2024ApJ...967..142T,2024ApJ...970..128S}), in tension with \rr{the theoretical predictions discussed above}. 

As an alternative to the (low-metallicity) SFR, long gamma-ray bursts (LGRBs), which are commonly associated with the collapse of massive stars into black holes, may better trace BBH progenitors. 
\rr{Even if LGRBs are not the direct progenitors of BBH mergers, they may indirectly trace BBH progenitors by tracing the birth of BHs more generally.}
Observationally, LGRBs are associated with sub-solar metallicity environments~\citep{2015A&A...581A.102V,2016ApJ...817....8P,2019A&A...623A..26P}.
LGRB jets are thought to be powered by the rotation of the collapsing stellar core \citep[``collapsar;"][]{1993ApJ...405..273W,1999ApJ...524..262M}.  
The mechanism by which massive stars maintain significant core rotation may be related to BBH evolutionary pathways.

Indeed, recent studies have proposed that the origin of (some) LGRBs is \rr{directly} linked to the formation of BBH mergers with spinning component BHs~\citep{2018A&A...616A..28Q,2022A&A...657L...8B}. These studies postulate that sufficiently rotating stellar cores are created from tight binaries (e.g. tidal spin-up in post common envelope systems or chemically homogeneous binary star evolution) that are also oftentimes the progenitors of BBH mergers~\citep{2000NewA....5..191B,2004MNRAS.348.1215I,2007A&A...465L..29C,2007Ap&SS.311..177V,2008A&A...484..831D}. 
Unless launching the LGRB spins down the BH~\citep{2024ApJ...961..212J}, the BBH system would also contain one (in the case of tidal spinup by a BH companion) or two (in the chemically homogeneous evolution case) spinning component BH in this scenario.
\citet{2022A&A...657L...8B} combine their population synthesis models with GW observations from GWTC-2 and LGRB observations from the SHOALS survey~\citep{2016ApJ...817....7P}, inferring that $\sim10\%$ of observed BBH mergers had an associated LGRB earlier in their evolution, and $20$--$85\%$ of LGRBs may have an associated BBH merger later in their evolution.
On the other hand, \citet{2022ApJ...933...17A} analyzed GW and LGRB observations and concluded that BBH mergers cannot come from LGRB progenitors unless they always experience long delay times greater than several Gyrs, \rr{whereas most predicted delay time distributions peak below 1 Gyr}. 

In this work, we revisit the \citet{2022ApJ...933...17A} analysis and compare the populations of LGRBs and BBH mergers and their evolution with redshift. Using the GWTC-3 observations, we infer the BBH delay time distribution \rr{under the hypothesis} that their progenitors follow the LGRB rate as a measured by \cite{2022ApJ...932...10G}, scaled by some factor. Over the currently observable range of BBH merger redshifts $0 < z < 1$, the BBH merger rate evolves slightly less steeply than the LGRB rate.
\rr{Thus, if BBH progenitor formation traces the LGRB rate, we find that the implied delay time distribution is perfectly consistent with the typical population synthesis prediction of a power law with slope $\alpha = -1$ and $\tau_\mathrm{min} = 10$ Myr.}
Contrary to the conclusions of \citet{2022ApJ...933...17A}, \rr{the consistency between our measured delay time distribution and its theoretical prediction supports the hypothesis} that LGRBs trace the BBH progenitor formation rate. 
However, \rr{this conclusion is only based on the shape of the redshift distributions of the two populations. Comparing the amplitude of the BBH merger rate and the LGRB rate, we find that} the LGRB rate is $\mathcal{O}(10)$ times higher than the BBH progenitor rate, so that BBH progenitors can make up only a small fraction of the LGRB rate. \rr{Thus, it is likely that LGRBs are indirectly related to the BBH progenitor population.} We also repeat our analysis under the assumption that only BBH mergers with significant aligned spins originate from LGRBs, in which case the delay time distribution may be steeper (favoring shorter delays).

\section{Methods}

\subsection{Applying delay time distributions to the LGRB rate}

For a given formation rate $\mathcal{R}_f$ and delay time distribution $p(\tau)$, the merger rate at a lookback time $t_L$ is given by:
\begin{equation}
    \label{eq:Rm}
    \mathcal{R}_m(t_L) = \int_{\tau_\mathrm{min}}^{ \tau_\mathrm{max}} \mathcal{R}_f(t_L + \tau)p(\tau) d\tau,
\end{equation}
where $\tau_\mathrm{min}$ and $\tau_\mathrm{max}$ are the minimum and maximum delay times, respectively. 
We assume that the maximum redshift of progenitor formation is $z = 15$, corresponding to a lookback time of $13.519$ Gyr, which we also fix as the maximum delay time,  $\tau_\mathrm{max} = 13.519$ Gyr.
We use the best fit cosmological parameters from~\citealt{2020A&A...641A...6P} throughout, as implemented by \textsc{Astropy}~\citep{2013A&A...558A..33A,2022ApJ...935..167A}.
The BBH delay time distribution is often modeled as a power law with power-law slope $\alpha$:

\begin{equation}
    p(\tau | \alpha) = \tau^\alpha \cdot \frac{\alpha + 1}{\tau_\mathrm{max}^{\alpha + 1} - \tau_\mathrm{min}^{\alpha + 1}} 
\end{equation}

\begin{figure}
    \centering
    \includegraphics[width=1\linewidth]{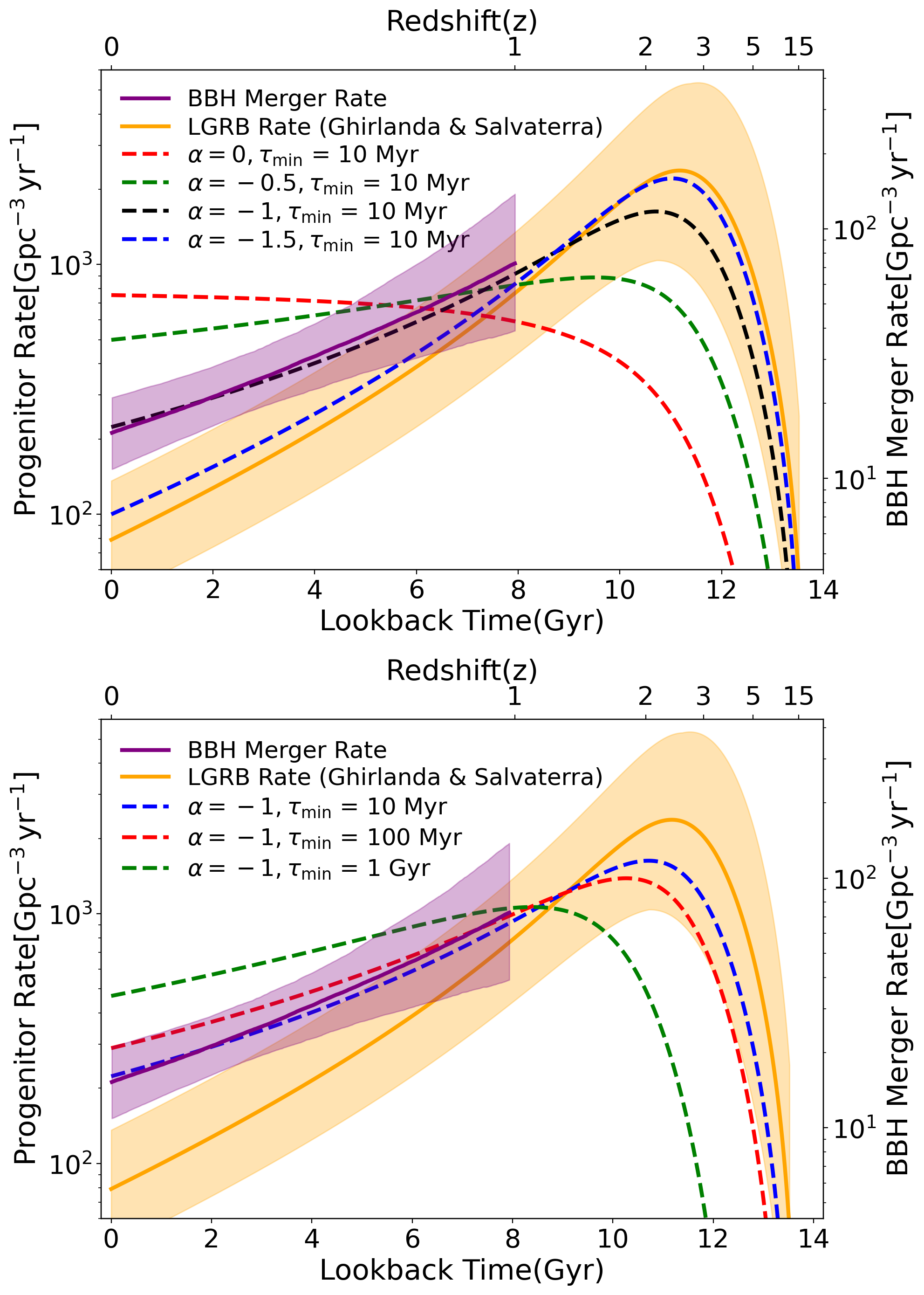}
    \caption{\label{fig:Rf-Rm-v-time} Upper panel: In orange (left axis), the maximum likelihood LGRB rate and its uncertainty from \cite{2022ApJ...932...10G}. The LGRB rate peaks at redshift $z = 2.6$, which corresponds to a lookback time $t_L = 11.2$ Gyr. Red, green, black and blue dashed curves show the maximum likelihood value of  $\mathcal{R}_{\rm LGRB}$ (solid orange line) convolved with power-law delay time distributions with slopes $\alpha = 0, -0.5, -1 , -1.5$ and $\tau_\mathrm{min} = 10$ Myr. At $\alpha = -2$, the merger rate is indistinguishable from the LGRB rate on this plot. In purple (right axis) is the BBH merger rate inferred by~\citet{2022A&A...665A..59B} up to $z = 1$, which is similar in shape to the $\alpha = - 1$, $\tau_\mathrm{min} = 10$ Myr curve (black dashed line). The right axis, corresponding to the measured BBH merger rate in purple, is scaled by a factor of around 0.06 relative to the left axis, corresponding to all other lines. Lower panel: Same as above, fixing $\alpha = -1$ and taking different values of $\tau_\mathrm{min}$ from 10 Myr (blue), 100 Myr (red) and 1 Gyr (green).} 
\end{figure}

In Fig.~\ref{fig:Rf-Rm-v-time}, we show some examples of the merger rate corresponding to a progenitor formation rate given by the best-fitting LGRB rate $\mathcal{R}_f \equiv \mathcal{R}_\mathrm{LGRB}$ from \citet{2022ApJ...932...10G} convolved with a few different power-law delay time distributions varying $\alpha$ (top panel) and $\tau_\mathrm{min}$ (bottom panel).
\rr{These curves correspond to the left y-axis labeled ``Progenitor Rate."}
The LGRB rate is shown in orange, with shaded bands denoting 68\% uncertainty.
For the uncertainty on $\mathcal{R}_\mathrm{LGRB}$, we use the reported one-dimensional (marginal) uncertainties on the inferred LGRB rate parameters from~\citet{2022ApJ...932...10G}, but we neglect correlations between the parameters, so our uncertainty bands are more conservative than the results of \citet{2022ApJ...932...10G}.
Delay time distributions with more negative $\alpha$ or smaller $\tau_\mathrm{min}$ favor shorter delay times, yielding predicted merger rates that are closer to the LGRB rate.
\rr{For example, the merger rate under the steepest delay time distribution we show, with $\alpha = -1.5$ and $\tau_\mathrm{min} = 10$ Myr (dashed blue line, top panel), approaches the best-fit LGRB rate (solid orange line).}
Any delay time distribution with $\alpha \lesssim -2$ and $\tau_\mathrm{min} = 10$ Myr would produce a merger rate that is indistinguishable from the LGRB rate on this plot.


We compare these predicted merger rates to the measured BBH merger rate as a function of redshift, as inferred by~\citet{2022A&A...665A..59B} from GWTC-3 observations.
\citet{2022A&A...665A..59B} perform a hierarchical Bayesian fit to the BBH mass, spin and redshift distributions, fitting a~\citet{2014ARA&A..52..415M}-like model to the BBH rate as a function of redshift while also allowing the BBH spin distribution to evolve with redshift. 
This measured BBH merger rate is shown in Fig.~\ref{fig:Rf-Rm-v-time} with the purple line denoting the median and shaded bands denoting 68\% credible intervals (right axis). 
Note that the right axis (BBH merger rate) is scaled by a constant factor of $0.06$ with respect to the left axis (corresponding to all other curves) for ease of comparison.

\rr{For the following analysis,} we define the constant scaling factor $f$ between the BBH progenitor formation rate and the LGRB rate:
\begin{equation}
\label{eq:f}
     f \equiv \mathcal{R}_f / \mathcal{R}_\mathrm{LGRB}.
\end{equation}
This scaling factor accounts for the possibility that not all LGRBs evolve into merging BBH.
Additionally, there may be BBH mergers that did not evolve from LGRBs (so that $f$ may be larger than 1).
Eq.~\ref{eq:f} simply assumes that the ratio between the BBH progenitor formation rate and the LGRB rate is constant across redshift.
\rr{The purpose of our study is to test the validity of this assumption by checking its implications for the delay time distribution against theoretical expectations.}

Looking at Fig.~\ref{fig:Rf-Rm-v-time}, we can see that the LGRB rate with a power-law delay time distribution given by $\alpha = -1$ and $\tau_\mathrm{min} = 10$ Myr \rr{(black dashed curve, top panel, or blue curve, bottom panel)}, scaled by a factor $f \approx 0.06$, matches the BBH merger rate well.
\rr{Such a delay distribution matches the theoretical predictions discussed in \S~\ref{sec:intro} remarkably well.}
In the following, we make this comparison statistically rigorous. 

\subsection{Inferring the delay time distribution parameters}


Our goal is to infer the delay time distribution parameters $\alpha$, $\tau_\mathrm{min}$ and the scaling factor $f$.
For each combination of $\alpha$, $\tau_\mathrm{min}$, $f$, and LGRB rate parameters $\theta_{\rm LGRB}$, we calculate $\mathcal{R}_m(t_L | \alpha, \tau_\mathrm{min}, f, \theta_{\rm LGRB})$ by numerically integrating Eq.~\ref{eq:Rm} on a grid with step size $d\tau = 1$ Myr.
The $\theta_{\rm LGRB}$ parameters, which define $\mathcal{R}_\mathrm{LGRB}$, are measured by \citet{2022ApJ...932...10G} and marginalized over in our analysis.


\begin{figure}
    \centering  \includegraphics[width=1\linewidth]{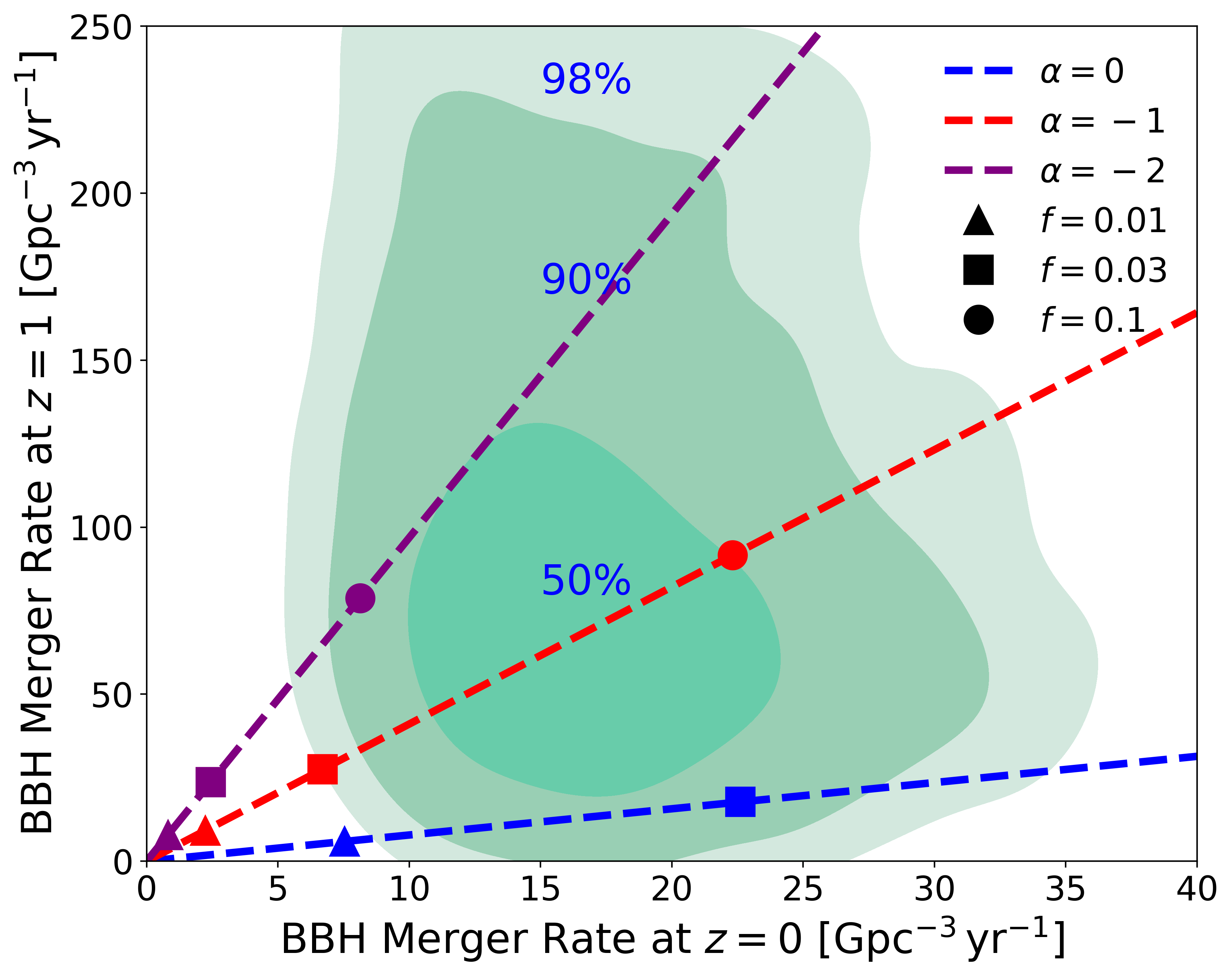}
    \caption{\label{fig:BBH-merger-rate-2d} The BBH merger rate at redshift $z = 0$ and $z = 1$, as inferred by \citet{2022A&A...665A..59B} and estimated here with a 2D KDE (green). Contours enclose regions of 50\%, 90\% and 98\% probability. Blue, red and purple dashed lines show predictions for the BBH merger rate under different power-law delay time distributions with slope $\alpha$ and $\tau_\mathrm{min} = 10$ Myr, assuming that the progenitor formation rate traces the best-fit LGRB rate from \citet{2022ApJ...932...10G}. A delay time distribution with $\alpha = -1$ (red)  crosses the high probability region. The circles, squares and triangles represent $f$ values (the ratio between the BBH formation rate and the LGRB rate) of 0.1, 0.03 and 0.01.} 
\end{figure}

We then quantify the similarity between any given $\mathcal{R}_m(t_L | \alpha, \tau_\mathrm{min}, f, \theta_\mathrm{LGRB})$ curve and the measured BBH merger rate $\mathcal{R}_\mathrm{BBH}(t_L)$ following the method in~\citet{2021ApJ...914L..30F}. 
We first take the BBH merger rate at $z = 0$ and $z = 1$ (the highest merger redshift probed by GWTC-3) from the population fit of \citet{2022A&A...665A..59B} and apply a 2-dimensional Kernel Density Estimation (KDE).
This is illustrated in Fig.~\ref{fig:BBH-merger-rate-2d}, which shows the 2-dimensional KDE of the BBH merger rate at $z = 0$ and $z = 1$, overlaid with example $\mathcal{R}_m$ values for various $\alpha$ and $f$ (fixing $\tau_\mathrm{min} = 10$ Myr and $\theta_\mathrm{LGRB}$ to the maximum likelihood value from \citealt{2022ApJ...932...10G}). 
Steeper delay time distributions (more negative $\alpha$) correspond to a merger rate that increases faster with increasing redshift, i.e. larger ratios $\mathcal{R}_\mathrm{BBH}(z = 1) / \mathcal{R}_\mathrm{BBH}(z = 0)$. 
Curves of constant $\alpha$ correspond to constant $\mathcal{R}_\mathrm{BBH}(z = 1) / \mathcal{R}_\mathrm{BBH}(z = 0)$ and appear as lines in Fig.~\ref{fig:BBH-merger-rate-2d}. 
From Fig.~\ref{fig:BBH-merger-rate-2d}, we see that $\alpha = -1$ is favored by the GW data because it passes through the central 50\% probability region of the KDE.
Meanwhile, increasing $f$ increases the merger rate at all redshifts, corresponding to points further away from the origin in Fig.~\ref{fig:BBH-merger-rate-2d}. 

Following \citet{2021ApJ...914L..30F}, we calculate a likelihood for any combination of delay time distribution and progenitor formation rate parameters by evaluating the 2-dimensional BBH merger rate KDE of Fig.~\ref{fig:BBH-merger-rate-2d} at the corresponding $\mathcal{R}_m(z = 0 | \alpha, \tau_\mathrm{min}, f, \theta_\mathrm{LGRB})$ and $\mathcal{R}_m(z = 1| \alpha, \tau_\mathrm{min}, f, \theta_\mathrm{LGRB})$ values.
We evaluate this likelihood on a grid of $\alpha$, \rr{ $\tau_\mathrm{min}$}, and $f$ values.
\rr{We also show results for a fixed value of} $\tau_\mathrm{min}$ = 10 Myr (the typical minimum delay time predicted from population synthesis because it roughly corresponds to the lifetime of massive stars\footnote{The delay time distributions predicted by population synthesis usually refer to the time between the birth of the progenitor star and the merger of the BBH. Here, we consider delay times between the birth of a BH (produced by the LGRB) and its merger in a BBH. This is shorter by $\sim10$ Myr (the typical massive star lifetime) than the traditional delay time distribution, which is too short a timescale to noticeably affect our results.}).
We marginalize over the uncertainty in the LGRB rate inferred by \citet{2022ApJ...932...10G} with a Monte Carlo average over 100 $\theta_\mathrm{LGRB}$ samples.
Because we do not have access to the full posterior distributions, we draw $\theta_\mathrm{LGRB}$ from the 1-dimensional credible regions reported by~\citet{2022ApJ...932...10G}, neglecting multi-dimensional correlations.
This is a conservative estimate of the uncertainty in $\mathcal{R}_{\rm LGRB}$.

From the likelihood over $\alpha$, \rr{$\tau_\mathrm{min}$}, and $f$, we calculate the posterior according to Bayes' theorem, choosing a flat prior on $\alpha$ and $\tau_\mathrm{min}$ and a flat-in-log prior on $f$. 
We present the resulting posterior on $\alpha$, \rr{ $\tau_\mathrm{min}$}, and $f$ in the following section.


\section{Results}
\begin{figure}
    \centering
    \includegraphics[width=0.5\textwidth]{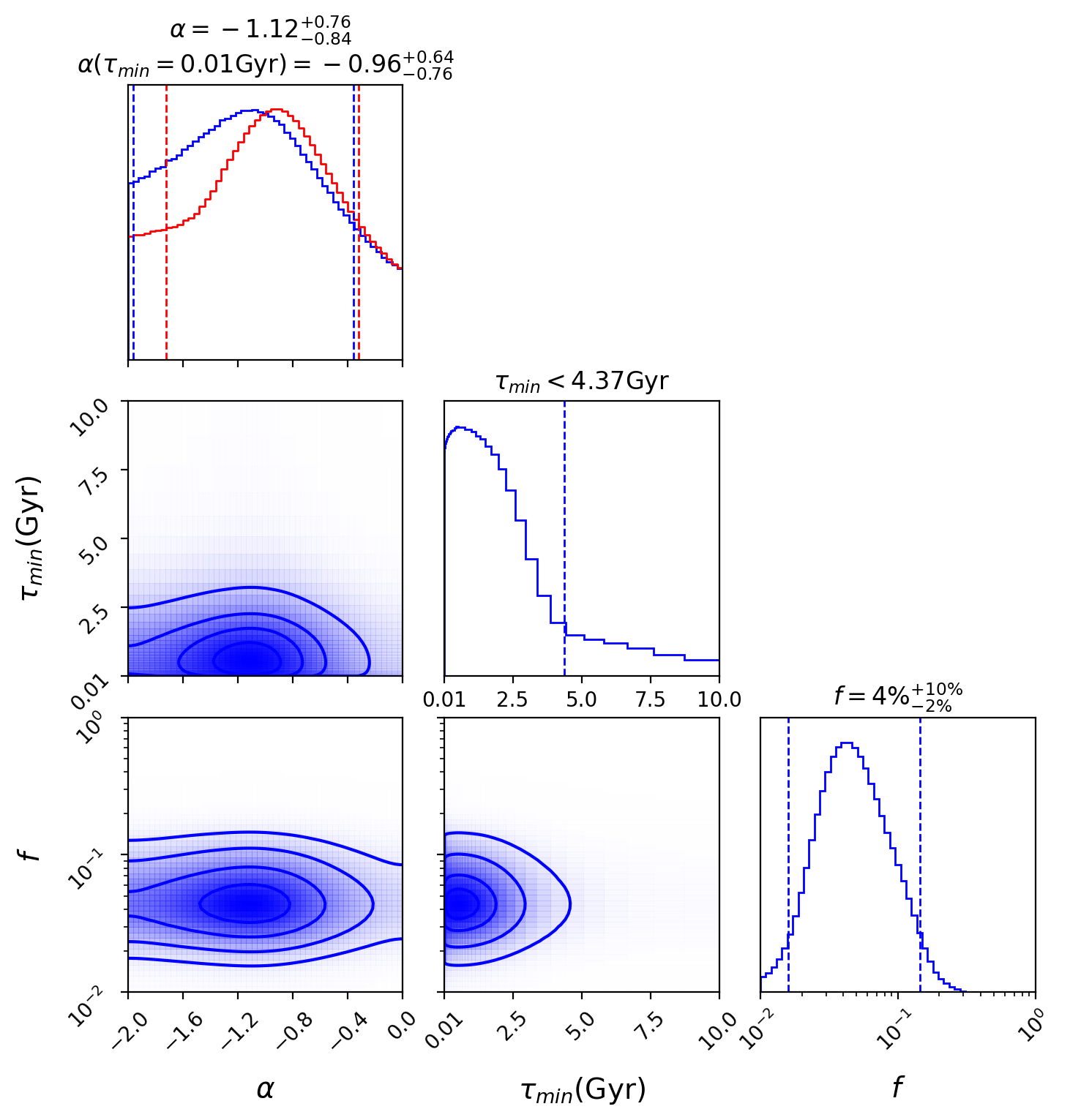}
    \caption{\label{fig:alpha-f-posterior-allchieff} Joint posterior on BBH delay time and formation rate parameters $\alpha$, $\tau_\mathrm{min}$ and $f$. Contours enclose 0.5, 1, 1.5 and 2 sigma posterior probability regions. Top Left: Posterior probability density for $\alpha$, marginalizing over $f$ and $\tau_\mathrm{min}$ (blue curve, with blue dashed lines denoting 90\% HPDI interval) or fixing $\tau_{\rm min} = 10$ Myr (red curve, with red dashed lines denoting 90\% HDPI interval). Center: Posterior probability density for $\tau_{\rm min}$, marginalizing over $\alpha$ and $f$ (blue curve, with dashed lines denoting 90\% upper bound). Bottom right: Posterior probability density for $f$, marginalizing over $\alpha$ and $\tau_\mathrm{min}$ (blue curve, with blue dashed lines denoting 90\% HDPI interval). We assume flat priors over $\alpha$ and $\tau_\mathrm{min}$ and a flat-in-log prior on $f$.}
\end{figure}

Figure~\ref{fig:alpha-f-posterior-allchieff} shows the joint posterior over the delay time distribution power-law slope $\alpha$, \rr{minimum delay time $\tau_\mathrm{min}$}, and the relative amplitude of the progenitor formation rate $f$.
We summarize our measurements with 90\% highest density posterior intervals (HDPI). 
Consistent with the results suggested by Fig.~\ref{fig:Rf-Rm-v-time} and Fig.~\ref{fig:BBH-merger-rate-2d}, we find $\alpha = -1.12^{+0.76}_{-0.84}$ (marginalizing over $f$ and $\tau_\mathrm{min}$) \rr{and $\alpha = -0.96^{+0.64}_{-0.76}$ (fixing $\tau_\mathrm{min} = 10$ Myr).}  
In other words, assuming a progenitor formation rate proportional to the LGRB rate, the GWTC-3 data prefer a nearly flat-in-log delay time distribution ($\alpha \approx -1$). 
As seen by the joint posterior between $\alpha$ and $\tau_\mathrm{min}$, larger $\tau_\mathrm{min}$ imply steeper power-law slopes, because the delay time distribution compensates to keep \rr{roughly} the same average delay time.
Marginalizing over $\alpha$, it is hard to distinguish between values of $\tau_\mathrm{min}$ as long as $\tau_\mathrm{min} \lesssim 1$ Gyr. Minimum delay times larger than 4 Gyr are ruled out \rr{at $>90\%$ credibility}, because they would imply a BBH merger rate that peaks at $z < 1$, which is not consistent with the GW data. 
Marginalizing over the delay time distribution, the BBH progenitor formation rate is likely $f = 4^{+10}_{-2}\%$ of the LGRB rate.


Next, we consider the scenario proposed by \citet{2022A&A...657L...8B} in which LGRBs only give rise to the subset of BBH mergers with moderately large spins that are aligned with the orbital angular momentum. 
Specifically, \citet{2022A&A...657L...8B} consider BBH mergers with effective inspiral spin $\chi_\mathrm{eff} > 0.2$. 
The fraction of BBH mergers with $\chi_\mathrm{eff} > 0.2$ is small, but may increase with increasing redshift~\citep{2022ApJ...932L..19B,2022A&A...665A..59B}. 
The population inference of~\citet{2022A&A...665A..59B} provides a measurement of the rate of BBH mergers with $\chi_\mathrm{eff} > 0.2$ as a function of redshift, \rr{allowing the spin distribution to evolve with redshift}. 
We repeat our analysis using this ``$\chi_\mathrm{eff} > 0.2$" BBH merger rate instead of the total BBH merger rate.

\begin{figure}
    \centering
    \includegraphics[width=0.5\textwidth]{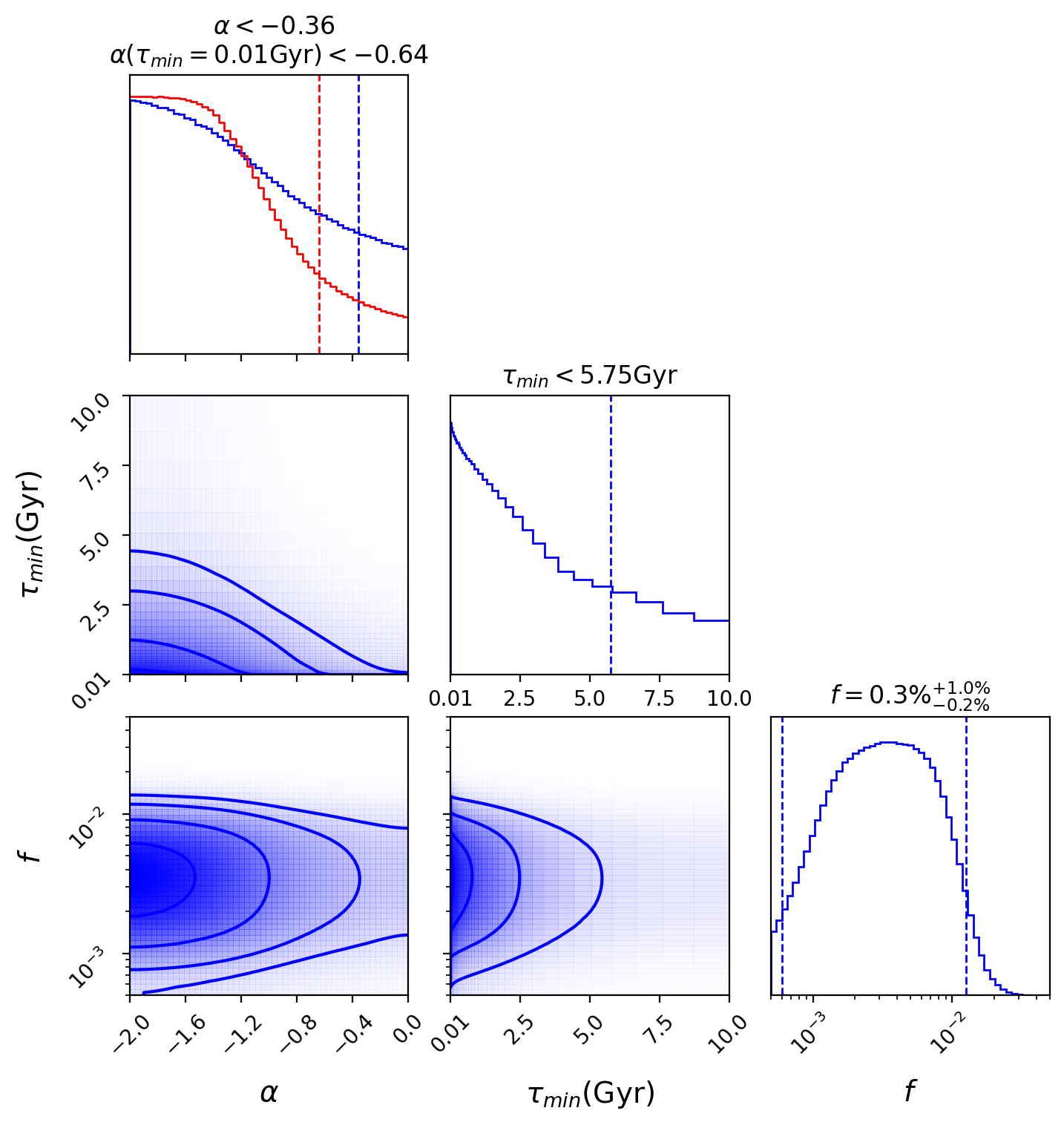}
    \caption{\label{fig:alpha-f-posterior-highchieff}
    Same as Fig.~\ref{fig:alpha-f-posterior-allchieff}, but using the subpopulation of BBH mergers with $\chi_\mathrm{eff} > 0.2$ rather than the full BBH population, assuming that LGRBs produce spinning BHs. 
    Compared to Fig.~\ref{fig:alpha-f-posterior-allchieff}, smaller values of $f$ and more negative values of $\alpha$ are preferred in this case, and large values of $\tau_\mathrm{min}$ are still disfavored. The rate of BBH mergers with $\chi_\mathrm{eff} > 0.2$ over the measured range $0 < z < 1$ is consistent with tracing the best-fit LGRB rate, so we infer steep delay time distributions peaking at our prior boundary $\alpha = -2$.}
\end{figure}

The resulting inference on $\alpha$, \rr{$\tau_\mathrm{min}$} and $f$ is shown in Fig.~\ref{fig:alpha-f-posterior-highchieff}. 
Under the assumption that only these high-spinning BBH mergers come from LGRBs, we estimate that their progenitor formation rate is most likely $<1\%$ of the LGRB rate. 
Meanwhile, the delay time distribution remains consistent with a flat-in-log distribution, but favors steeper power-law slopes, railing against the prior boundary at $\alpha = -2$ (recall that steeper delay time distributions are virtually indistinguishable from $\alpha = -2$). 
This is because the slope of the high-spin BBH merger rate agrees with the slope of the LGRB rate from redshift 0 to 1, requiring essentially no delay times.
Within our prior boundaries, we constrain $\alpha < -0.36$ (marginalizing over $\tau_\mathrm{min}$) or $\alpha < -0.64$ (fixing $\tau_\mathrm{min} = 10$ Myr) at 90\% credibility.
This may imply that BBH mergers with high aligned spins experience shorter delay times than slowly-spinning BBH systems, which is consistent with predictions that spinning BBHs originate in tighter binary orbits that merge faster~\citep[e.g.][]{2022A&A...665A..59B}.

\section{Discussion and conclusions}

By comparing the redshift evolution of LGRBs and BBH mergers, we find that the BBH merger rate is consistent with tracing a scaled LGRB rate with a flat-in-log delay time distribution (with a best fit power-law slope of $\alpha \approx -1$, under which 60\% of BBH mergers occur within 1 Gyr of formation). 
This delay time distribution is consistent with theoretical \rr{population synthesis} predictions for the formation of BBH mergers from several channels, including isolated binary evolution and dynamical assembly in dense star clusters\footnote{For BBH assembly in star clusters, the delay time is defined as the time between cluster formation and BBH merger.}, \rr{many of which predate GW observations}~\citep{2010ApJ...716..615O,2012ApJ...759...52D,2017MNRAS.472.2422M,2020MNRAS.498..495D,2023ApJ...957L..31F,2024ApJ...967...62Y}.
Previous work has assumed that the BBH progenitor formation rate traces the SFR (often with some metallicity dependence), in which case significantly steeper power-law slopes are preferred.
The consistency between our recovered delay time distribution and theoretical predictions suggests that the LGRB rate is a better tracer of the BBH progenitor formation rate than the SFR, at least under the previously used assumptions for the metallicity-specific SFR and the metallicity dependence of BBH progenitors.
\rr{In other words, if we instead performed the ``backwards" analysis of~\citet{2023ApJ...957L..31F} and applied their theoretically-motivated delay time distribution to the observed merger rate (their fiducial distribution had $\alpha = -1$ and $\tau_\mathrm{min} = 10$ Myr), we would find that the recovered BBH progenitor formation rate is proportional to the LGRB rate.}

Nevertheless, our results do not definitively prove that LGRBs are progenitors to BBH mergers. We find that BBH progenitors are much rarer than LGRBs, accounting for only $4^{+10}_{-2}\%$ of the LGRB rate.
This is smaller although not entirely inconsistent with \citet{2022A&A...657L...8B}, who report that 20--85\% of LGRBs may be produced by the progenitors of BBH systems.   
However, if LGRBs produce spinning BHs, we find that this fraction is much smaller: $0.3^{+1.1}_{-0.2}\%$.
Nevertheless, \citet{2023ApJ...952L..32G} and \citet{2024ApJ...961..212J} have recently argued that only moderate rotation of the stellar core is necessary to power an LGRB, and that the LGRB itself may spin down the resulting BH. 
The fact that so few BBH mergers have large aligned spins may lend support to their argument.

Rather than a direct relationship between LGRBs and BBH progenitors, it may instead be that both populations trace the low-metallicity SFR with similar metallicity thresholds. 
In this case, it is still convenient to use LGRBs as probes of the low-metallicity SFR, because they are detected out to high redshifts, whereas low-metallicity SFR occurs mainly in small, faint galaxies that are hard to directly observe.
Rather than direct measurements of the SFR in these galaxies, \citet{2022ApJ...937L..27M} proposed cross-correlating line intensity maps with BBH mergers in order to constrain the BBH delay time distribution, a promising technique once line-intensity mapping surveys reach completion. Future \rr{GW observations, combined with measurements of the LGRB rate and deeper observations of the low-metallicity, high redshift SFR} would enable us to better understand the metallicity-dependence of the LGRB and BBH populations, with implications for stellar collapse, supernova explosions and binary evolution.

\begin{acknowledgments}
We thank Michael Zevin for helpful comments on the manuscript. 
MF is supported by the Natural Sciences and Engineering Research Council of Canada (NSERC) under grant RGPIN-2023-05511. 
MF is grateful to the Lorentz Center, the scientific organizers and the participants of the workshop ``Gravitational waves: a new ear on
the chemistry of galaxies" (29 April - 3 May 2024,
\url{https://www.lorentzcenter.nl/gravitational-waves-a-new-ear-on-the-chemistry-of-galaxies.html}) for helpful discussions.  
This material is based upon work supported by NSF's LIGO Laboratory which is a major facility fully funded by the National Science Foundation.
\end{acknowledgments}

\bibliography{references}{}
\bibliographystyle{aasjournal}

\end{document}